\documentclass[aip,amsmath,amssymb,preprint]{revtex4-1}

\usepackage[english]{babel}
\usepackage{graphicx}
\usepackage{dcolumn}
\usepackage{bm}
\usepackage{CJK}
\usepackage{siunitx}
\usepackage[utf8]{inputenc}
\usepackage[T1]{fontenc}
\usepackage{mathptmx,amsmath,amssymb}

\begin{document}


\title{Non volatile photo-switch using a diamond pn junction }

\author{C\'edric Masante} 
\affiliation{Univ. Grenoble Alpes, CNRS, Institut N\'eel, 38000 Grenoble, France} %

\author{Martin Kah} 
\affiliation{Univ. Grenoble Alpes, CNRS, Institut N\'eel, 38000 Grenoble, France} %

\author{Cl\'ement H\'ebert} 
\affiliation{Univ. Grenoble Alpes, INSERM, U1213, Grenoble Institut des Neurosciences (GIN), Grenoble, France} %

\author{Nicolas Rouger} 
\affiliation{LAPLACE, Universit\'e de Toulouse, CNRS, INPT, UPS, Toulouse, France} %

\author{Julien Pernot}\thanks{julien.pernot@neel.cnrs.fr}
\affiliation{Univ. Grenoble Alpes, CNRS, Institut N\'eel, 38000 Grenoble, France} %

\begin{abstract}

Ultrawide bandgap semiconductor technologies offer potentially revolutionary advances in the rapidly developing areas of quantum communication, short wavelength optics, smart energy conversion and biomedical interfaces. These strongly demanding technologies can be partly constructed using conventional devices but new hybrid architectures are needed to overpass current performances and add functionalities. Here, we propose a new concept based on the specific properties of a diamond pn junction combined with both an electric and optical control of the depletion region. Using this junction as a gate in a junction field effect transistor, we report a proof of concept of a non volatile diamond photo-switch. A diamond pn junction made with nitrogen deep donors in the n?side is demonstrated to be optically activated thanks to visible light. The n-type diamond gate is almost devoid of free carriers in the dark and thus insulating. Illuminating the device renders the standard electrical gate control of the transistor efficient. Without illumination, the device is frozen, keeping a permanent memory of the current state. This new way of operating the device opens numerous possibilities to store and transfer information or energy with applications in the field of electrical aircraft or aerospace electronics, power electronics, bio-electronics and quantum communication. 

\end{abstract}


\maketitle

\section{Introduction}

Because of their exceptional characteristics, ultrawide bandgap semiconductors (UWBGS) like diamond, aluminium nitride, boron nitride, gallium oxide are already in focus for producing devices beyond state-of-the-art semiconductor technologies \cite{Tsao2018}. For electronics, most researchers and engineers are still using concepts envisaged for silicon and wide bandgap semiconductors (like silicon carbide and gallium nitride) where much smaller band gap energies are at stake. The limits of this incremental approach for the fabrication of UWBGS-based devices are already well known and established. In this paper, we propose to go beyond current paradigms by putting forward the design of novel hybrid technologies (optical and electrical) based on unconventional concepts which exploit the huge energy scale of UWBGS.\\

Diamond's outstanding properties like wide bandgap, elevated carrier mobility, high breakdown field and huge thermal conductivity make it a unique material for high power and/or high temperature and/or radiation-resistant electronics. The long valley-polarization relaxation times of electrons make diamond a good candidate for use as valleytronics transistor \cite{IsbergNatMater2013,Suntornwipat2021}. Diamond takes also advantage of point defects, like nitrogen vacancy centres, for applications in quantum sensing \cite{Maminscience2013}, quantum communication \cite{HensenNature2015} or spin-based photonic quantum technologies \cite{AtatureNatureReviews2018}. Finally, its bio-compatibility offers the possibility of integrating diamond electronics into the human body \cite{Rifai2021,Lichter2015,Fox2020,Mani2020,Fan2020,Dankerl2009,Piret2015}. To summarize, this material provides new opportunities to locally probe electric or magnetic signals, to reliably process data and to safely communicate them in harsh or living environments. Diamond devices capable of storing charges in a non volatile manner or addressing individual devices are needed to develop monolithic diamond-based systems. As examples, the charge accumulations could be used for power supply (local on-chip energy storage) or information storage with read and write access (non volatile memory) and individual on-chip addressing for application in neuronal multiplexers. Embedded on-chip devices coupled with an ultra high conversion efficiency of beta-voltaic cell using diamond $pn$ junction \cite{ShimaokaAPL2020} can be envisaged for aerospace or harsh environment applications.\\

In this paper, we report on a diamond-based non volatile photo-switch with a permanent off state having no loss. The concept is demonstrated by using monocristalline $p-$type diamond deposited by plasma enhanced chemical vapour deposition on a highly nitrogen-doped diamond substrate manufactured using high pressure and high temperature. In the first part, the concept will be described. Then, the active and freeze state of the $pn$ junction will be established and the proof of concept of a non volatile photo-switch will be demonstrated thanks to an electro-optical control of a diamond junction field effect transistor. Then, we will discuss examples of application fields in which this concept can efficiently operate such as power electronics, quantum communication or bio-electronics. 

\section{Non volatile diamond-based photo-switch concept}

For analogue or digital operations, the ubiquitous silicon-based field effect transistors work by using a gate bias voltage to deplete (OFF state) or populate (ON state) the channel. In the hybrid structure proposed here, the external illumination activates or freezes the device. In this way, the device can be driven by a combination of electrical and optical signals, providing a higher degree of freedom. In conventional semiconductors with shallow energy levels associated to donors or acceptors, doping level and carrier concentrations are almost equal at room temperature. The doped material is conducting. In nitrogen doped diamond, nitrogen impurities in substitutional sites create deep donor levels at 1.7 eV from the conduction band. Consequently, at room temperature, electrons remain bound to the dopants. Even at high doping levels (10$^{18}$ cm$^{-3}$ to 10$^{19}$ cm$^{-3}$) nitrogen doped (N-doped) diamond samples are completely insulating, and so cannot be electrically driven. This is the \textit{freeze state} of the device. Thanks to external illumination, the absorption of photons promotes some of the bound electrons to the conduction band, increasing the free carrier concentration. The device thus becomes conducting and prone to be electrically driven. This is the \textit{active state} of the device. The control of the ON and OFF states of the transistor is still ensured by a gate bias voltage. However, the switch from ON to OFF will only be achievable during the active state. This concept has to be distinguished from photo-conductive semiconductor switch devices \cite{majda-zdancewiczCurrentStatePhotoconductive2018} which have also been demonstrated to use N-doped diamonds \cite{hallPhotoconductiveSwitchHigh2020} well. Optically generated carriers most participate in conduction which means these devices require a high power pulsed laser source to be efficient. For example, the device reported in \cite{hallPhotoconductiveSwitchHigh2020} makes use of a pulsed laser with pulse power of about 6.7 kW (the average power is 25 mW). In this work, a bulk boron doped ($p-$type) diamond is used as the channel and the nitrogen doped diamond as the gate. The lower ionization energy of boron acceptor (0.38 eV) means it can conduct at room temperature in the dark while the gate requires external optical excitation to generate free electrons. Optically generated carriers do not significantly participate in the channel conduction. Only the n-type gate junction is optically controlled by the light source.

\section{Optically controlled diamond pn junction}
	
\begin{figure}[t]
\includegraphics[width=1\textwidth]{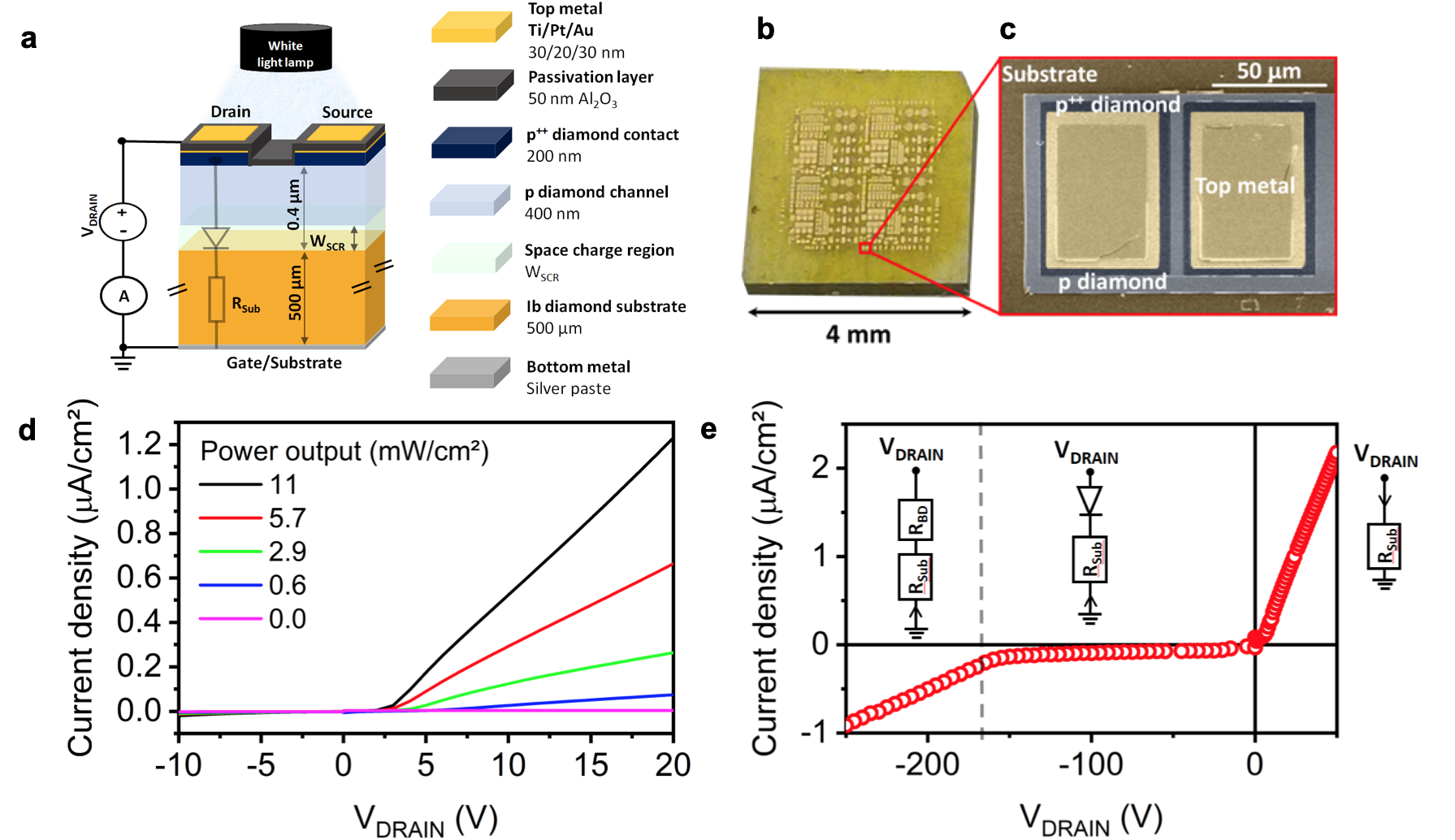}
\caption{\textbf{Optically controlled diamond pn junction} \textbf{a} Schematic representation of the device and the measurement setup during the $pn$ diode electrical characterization. The $p-$type layer resistance is neglected in comparison with the substrate resistance. \textbf{b} Optical view of the sample. \textbf{c } Secondary electron microscope top view image of a singular test device (colours added). \textbf{d} Room temperature current-voltage characteristic of the drain-gate $pn$ diode activated by a white light LED with various optical power ranging between 0 and 11 mW/cm?. The source is floating and the gate/substrate is grounded. The current is normalized by the $p-$type layer area delineated by mesa etching. \textbf{e} Current-voltage characteristic of the $pn$ diode under light exposure at $\SI{-50}{\celsius}$.}
\label{Figure1}
\end{figure}

A schematic description of the sample structure is shown in \textbf{Figure~\ref{Figure1}}\textbf{a} and top view images in \textbf{Figure~\ref{Figure1}}\textbf{b} and \textbf{\ref{Figure1}}\textbf{c}. A white LED driven by a tuneable current was used to optically excite the sample from the top surface. The average power density of the beam was measured with a photo-diode detector, ranging from 0 to 11 mW.cm$^{-2}$. Part of the light was absorbed and reflected by the top metallic layer. The \textbf{Figure~\ref{Figure1}}\textbf{d} shows a room temperature current-voltage characteristic under various optical power varying between 0 and 11 mW.cm$^{-2}$. A diode characteristic was clearly measured under optical excitation while the dark current was below detection limit (< 10$^{-8}$ A.cm$^{-2}$) in the whole voltage range. The threshold voltage was around 3 V, as expected for a diamond $pn$ junction where the Fermi level lies close to the deep nitrogen donor level in n-side region. This clearly demonstrates that a nitrogen doped and boron doped diamond junction can be optically activated by continuous low power white light excitation. The LED does not emit in the UV range required for electron-hole generation in diamond ($\lambda<225$ nm) \cite{KoizumiSceince2001}. Also, no intrinsic carriers were generated from the valence band to the conduction band. The linear relationship between the substrate conductivity and the irradiance (Figure~\ref{Figure1}\textbf{d}) is characteristic of a one photon impurity ionization process. In this experiment, the white light was found to be efficient to observe a photo-current because the transition from the nitrogen impurity level to the conduction band is efficient for wavelengths lower than 550 nm with absorption coefficients of the order of 10 to 100 cm$^{-1}$ \cite{deweerdtDeterminationDefectConcentration2008}. The corresponding absorption depths were between 1 and 0.1 mm, from the diamond surface. As expected, the forward current was greatly limited by the thick, resistive $n-$type substrate with a current density in the $\mu$A.cm$^{-2}$ range in the forward regime at room temperature (the current was normalized by the $p-$type layer surface area, delineated by the mesa etching). 
One can however extrapolate that the pn junction conductivity could be increased by orders of magnitude by (i) using a higher power continuous light sources up to 1000 W.cm$^{-2}$ light output power than can be achieved with LEDs or lasers, (ii) depositing transparent top contacts or illuminating from the backside of the sample, (iii) using a <500 nm light source to maximize the optical absorption, (iv) using a thinner N-doped substrate and (vi) forming a proper ohmic contact on the N-doped diamond. However this simple setup provides a strong proof of concept of the nitrogen-doped diamond electrical activation under light exposure.

\section{High breakdown capability of the pn junction gate}

As an UWBGS, diamond is expected to be able to sustain an extreme electric field before reaching its breakdown. Hiraiwa et al.\cite{hiraiwaFigureMeritDiamond2013} estimated the diamond breakdown field for a doping level [N$_{A}$]$=2\times 10^{17}$ cm$^{-3}$ at 10 MV.cm$^{-1}$ and Kamakura et al.\cite{kamakuraInitioStudyAvalanche2016} at 7 MV.cm$^{-1}$, compared to 0.2 to 0.6 MV.cm$^{-1}$ for silicon. However experimental demonstrations are challenging in diamond devices at the moment due to the presence of a strong leakage current which causes an early soft breakdown. To circumvent this difficulty in this original structure, we performed the measurement at $\SI{-50}{\celsius}$ in order to reduce the temperature dependent leakage current. Since the substrate carriers were optically generated, and the substrate was still more resistive than the $p-$layer, no significant variation of the diode resistance was measured. In the dark, the reverse bias is held by the full thickness of the insulating substrate and the current is below the detection limit for a bias up to 1 kV. However when the diode was reverse biased under optical excitation, because the N-doped substrate is activated with a doping level much higher than the $p-$type layer, the Space Charge Region (SCR) was found to extend almost exclusively in the $p-$side. Almost all the potential drop is then sustained by the 400 nm boron doped layer. The \textbf{Figure~\ref{Figure1}}\textbf{e} shows the obtained IV curve, exhibiting a clear change of slope at V=-156 V attributed to the junction's breakdown. Due to the large substrate serial resistance, the breakdown characteristic did not exhibit the sharp increase of current with a power law, but was instead immediately limited by the substrate resistance and appears linear. A local breakdown, not occurring on the full surface of the diode, might explain the higher serial resistance measured after the breakdown compared to the forward regime. This is represented by an additional serial resistance $R_{BD}$ in Figure~\ref{Figure1}e. The maximum electric field at breakdown has been estimated to be 5.2 MV.cm$^{-1}$ in a 1D punch-through configuration. This is a remarkably high value which is among the highest critical fields reported for diamond and much larger than other common wide band gap materials such as SiC and GaN (ideal breakdown around 2-3 MV.cm$^{-1}$). Since no special care to reduce the 2D and 3D effects was implemented in the fabrication process, the real value is most probably higher due to the well known electric field crowding effect. The high breakdown voltage of diamond pn junction will ensure the fabrication of reliable and robust devices able to work in high voltage operations under illumination but even more importantly, it ensures a galvanometric insulation of the source to the gate of the junction field effect transistor described below without illumination. 

\section{Electro-optical control of a diamond transistor}

\begin{figure}[t]
\includegraphics[width=1 \textwidth]{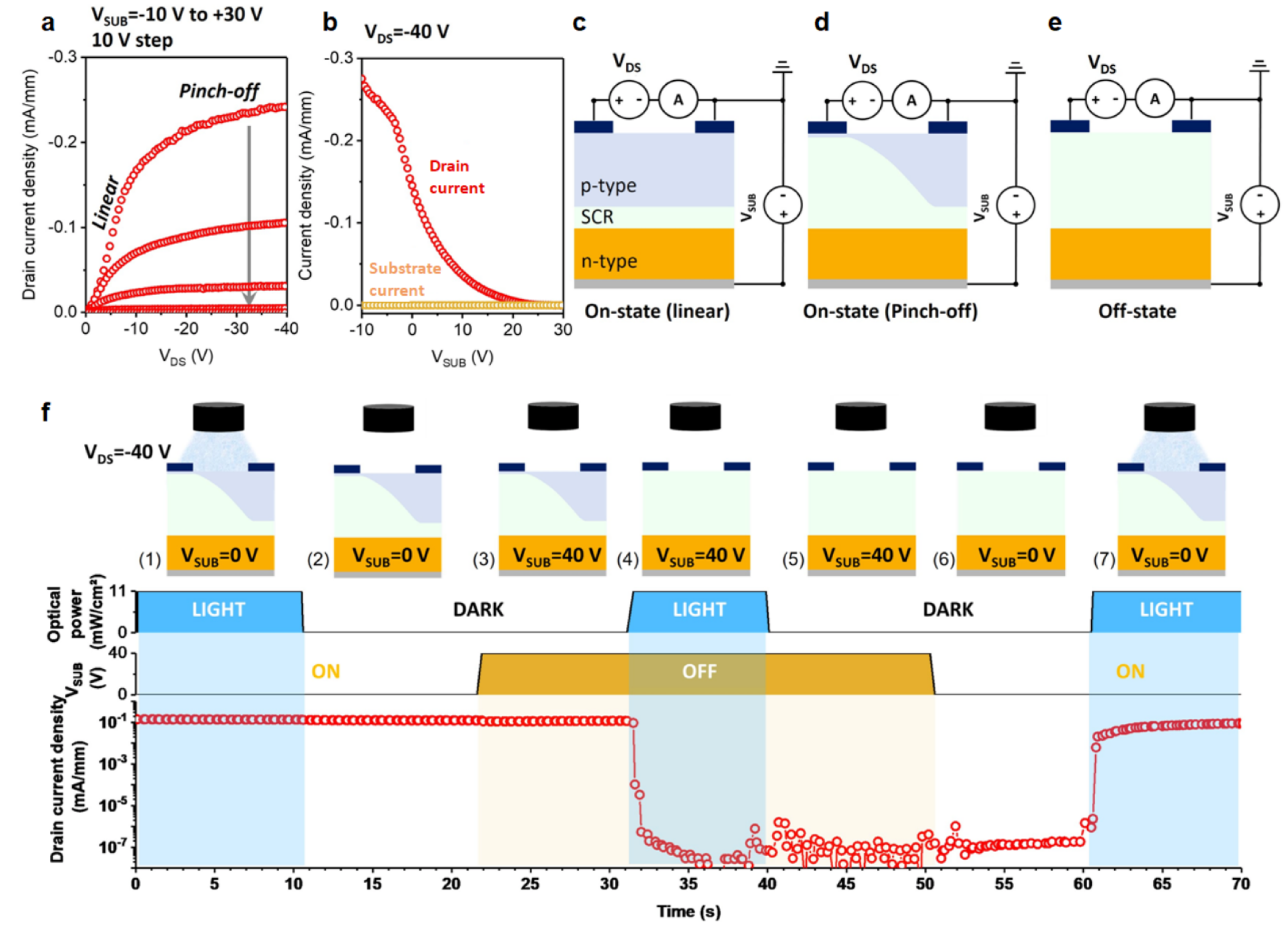}
\caption{\textbf{Non volatile photo-switch using a diamond junction field effect transistor.} \textbf{a} Quasi-static room temperature transistor characteristic, drain current versus drain-source voltage, measured for a substrate to source bias V$_{Sub}$ varying from -10 V to +30 V with 10 V step and continuous 11 mW/cm? optical power density.\textbf{ b} Drain current density as function of V$_{Sub}$ in the Pinch-off regime at V$_{DS}$=-40 V. \textbf{c} Schematic cross section of the device in ON-state at low V$_{DS}$ (linear regime), \textbf{d} at high V$_{DS}$ (pinch-off regime) and \textbf{e} in OFF-state. \textbf{f} Measured switched drain current over one commutation cycle of the fabricated JFET at $V_{DS}=-40$ V, as function of the light and substrate bias. The device is only able to commute under an optical signal.}
\label{Figure2}
\end{figure}

A junction field effect transistor device was fabricated on the same sample to demonstrate the first non volatile photo-switch using diamond \textit{pn} junction. The current flows through the bulk of the $p-$type layer between the drain and source under bias. Under optical illumination, the gate contact on the back side of the substrate controls the SCR extension in the channel. The drain to source distance is 10 $\mu$m and contacts have a 70 $\mu$m width. It is remarkable that room temperature transistor characteristics under an optical power of 11 mW/cm? were obtained as shown in \textbf{Figure~\ref{Figure2}}\textbf{a} and \textbf{\ref{Figure2}}\textbf{b}. The typical ON-state linear regime, pinch-off and OFF-state are schematically drawn in \textbf{Figure~\ref{Figure2}}\textbf{c}, \textbf{\ref{Figure2}}\textbf{d}, and \textbf{\ref{Figure2}}\textbf{e}, respectively. The pinch-off is due to the extension of the SCR close to the drain contact, when the drain to gate bias approaches or is greater than the threshold bias (i.e. the drain to source bias $V_{DS}$ is around -30 V or more for $V_{SUB}=0$ V). Due to the low carrier density in the nitrogen doped substrate, the gate junction can be forward biased (i.e. $V_{SUB}\lesssim -4$ V) without any significant gate current. The SCR modulated by the gate bias can be fully removed to increase the channel conduction (0.25 mA/mm). Figure~\ref{Figure2}a shows that a gate to source bias $V_{SUB}=-10$ V significantly increases the drain current. The gate current remains orders of magnitude lower, i.e. less than $2\times 10^{-6}$ mA.mm$^{-1}$. No drain current modulation from the gate could be observed without optical illumination.\\

A measured commutation cycle at a fixed drain to source bias V$_{DS}=-40 V$ (Pinch-off) is shown in \textbf{Figure~\ref{Figure2}}\textbf{f}, where the freezing of the device's electrostatic can be optically controlled. The following paragraph describes each step represented in Figure~\ref{Figure2}f: (1), the substrate to source bias V$_{Sub}$ is set at 0 V with the light on, a bulk conduction channel between the drain and source is obtained (normally-on device). (2), the light is switched off, the free electron concentration in the N-doped substrate is extremely low (\textit{freeze state}). In an ideal case without compensation from acceptor ions at room temperature, the free electron concentration is below $10^{4}$ $cm^{-3}$. (3) The gate substrate electric command was applied (V$_{Sub}=40 V$) to fully deplete the channel (OFF state). However, because the N-doped gate is devoid of free carriers in the dark, no electrostatic modulation can occur in the dark and the channel remains conductive even if the gate voltage corresponds to a OFF state. (4) The light is switched on, providing the optical signal activating the device (\textit{active state}). The device was able to respond to the electrical gate command and switches to OFF state. The channel was fully depleted and the drain current $I_{DS}$ dropped below the detection limit. (5) The light was switched off (\textit{freeze state}). (6) The gate electric command to switch the device in ON state was applied, V$_{Sub}=0$ V, but the channel remained insulating due to the absence of optical excitation. (7) The light was switched on (\textit{active state}) and the device was able to switch to the ON state.\\ 

\begin{figure}[ht!]
\includegraphics[width=1 \textwidth]{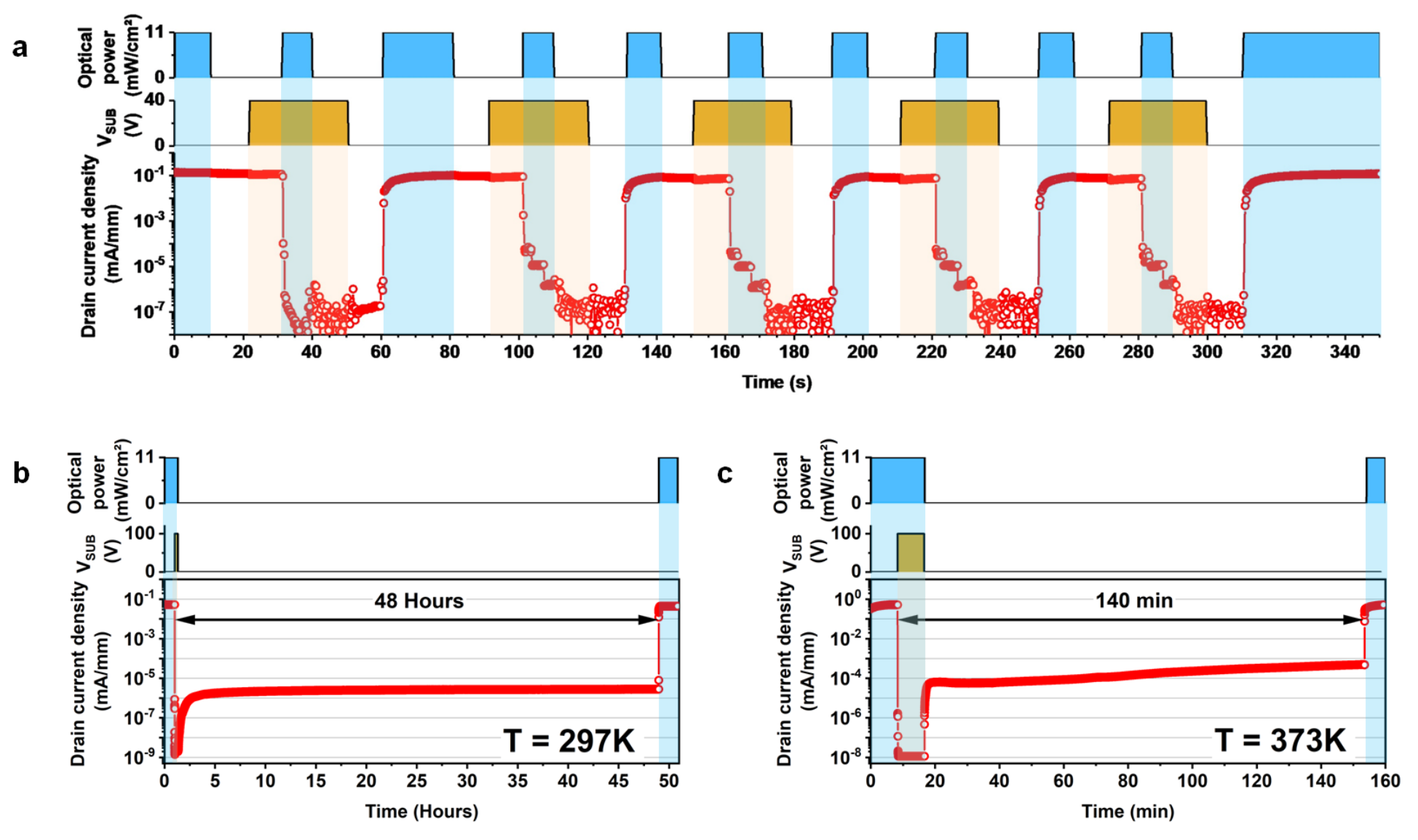}
\caption{\textbf{Temporal stability of the non-volatile photo-switch. } \textbf{a} Drain current density versus time for V$_{DS}$=-40 V at T=297 K for multiple commutation cycles alternating freeze/active and on/off states by using various substrate voltages and optical powers as described by the chronogram. Drain current versus time for V$_{DS}$=-40 V  \textbf{b} at T= 297 K and \textbf{c} T= 373 K , for long duration in order to illustrate the non volatile character of the off state in the freeze state.}
\label{Figure3}
\end{figure}

This sequence was replicated several times as shown in \textbf{Figure~\ref{Figure3}}\textbf{a} to ensure the reliability of the measurements. Although the first cycle seems different, all the following cycles were found to exhibit the same characteristic. In order to check the non volatile character of this optically activated switch, the device was tested during a long time in the off and freeze state. \textbf{Figure~\ref{Figure3}}\textbf{b} and \textbf{\ref{Figure3}}\textbf{c} shows the drain current versus time for V$_{DS}$=-40 V at room temperature and $\SI{100}{\celsius}$, respectively for long duration (2 days at room temperature and 140 minutes for 373 K). One can observe that the $I_{ON}/I_{OFF}$ ratio is larger than 4 orders of magnitude during more than 48 hours at room temperature and larger than 3 orders of magnitude during more than 1 hour at $\SI{100}{\celsius}$. This non-volatile photo-switch benefits from the unique properties of diamond such as its huge breakdown field and high temperature operation. \\

The large RC time constant and, more precisely, the huge resistance R caused by the high substrate's thickness and resistivity meant that the turn-on time was long but the modulation was clearly observed. Effects of traps most likely present at the substrate/epilayer and epilayer/oxide interfaces could not be distinguished. Reducing the turn-on time is a key factor in obtaining an efficient device, but as discussed previously, this could be decreased by orders of magnitude as it directly depends on the N-doped layer resistance. Additionally, in this kind of FET device the optical excitation only has to be present during the short time of the device switching. As well as the previously mentioned ways to increase this simple proof of concept performances, the use of a pulsed light source could provide a stronger optical excitation with limited total power usage.

\section{Non volatile photo-switch for new applications}

The next challenge to create new devices will be to grow nitride-based light emitting diodes (LED) directly on diamond transistor. Indeed, creating electrically driven visible or sub-band gap UV LEDs on top of the diamond transistors will enable high optical power density excitation to be attained. This would in turn enable the efficient control of the non volatile photo-switch, i.e. to read or write the transistor state (ON or OFF thanks to the FET gate) and store or reset the transistor state (freeze or activate thanks to the LED electrode). This new device paves the way for numerous practical applications where the superior properties of diamond semiconductors can be advantageously exploited. For embedded electronics, non volatile memories, i.e. the association of a switch with a capacitance for example, could be very useful to store and read information in a harsh environments (radiation resistant and high temperature) like in aerospace and aeronautic applications. For power electronics, the gate terminal of classical power transistors is affected by high $dV/dt$ and Miller coupling, causing undesired turn-ON and possible catastrophic shoot-through. Thanks to the perfect galvanic insulation of the gate terminal without illuminating light, the Miller coupling between the Drain and the Gate is completely removed thus offering the strongest immunity to parasitic turn ON. For quantum communication, the LED will make the excitation of a nitrogen vacancy (NV) centre possible but more importantly the diamond FET will also permit control of electrostatics and a non-volatile storage of the NV charge state. Finally, we shall describe a bio-electronics applications in which the unique properties of diamond in the context of this concept are fully exploited. Diamond-based electronic is a perfect candidate for chronic biomedical implants since it provides solution for most of the scientific barriers that prevent most advanced technology to be transferred to clinical applications. Its chemical inertness combined with its carbon nature make it a biocompatible material compatible with all the standard medical imaging techniques including Magneto Resonance Imaging \cite{Hebert2015}. Furthermore, diamond electronic can be easily sterilized without damaging its performances. Diamond electrodes and transistors have thus already proved to be very interesting candidate combination for interaction with the body for both sensing and stimulation \cite{Dankerl2009,Piret2015}. This is particularly true in the case of neural implants where bidirectional communication is required to develop neural prosthesis, help neural regeneration and gain further understanding of brain signal processing and dynamics. Future diamond embedded electronics have the potential to create implants with a high level of stability that do not require heavy hermetic insulation as is the case with silicon and other emerging polymer and flexible logic electronics \cite{Mani2020}. The concept offered by the photo-switch using a diamond pn junction is one of the first diamond implantable logic electronics which can be combined with a diamond sensor to develop a low power diamond time division multiplexer. Each line of an array of solution gated field effect diamond transistor can be sequentially read as in the case of CCD sensor, allowing for the reduction of the cumbersome electric wires that dramatically increase the size of the implant \cite{Schaefer2020}. Additional the freezing state of the switch means low power consumption can be achieved which is a key parameter for chronic implants.

\section{Conclusion}
We demonstrated a novel concept of device with a hybrid optical and electric gate control. The $pn$ junction made of N-doped diamond can be optically controlled using a white light LED. A breakdown of -156 V at $\SI{-50}{\celsius}$ has been measured, corresponding to a peak electric field estimated to be superior to 5.2 MV.cm$^{-1}$. This proof of concept of hybrid electrical and optical switching device consists of a diamond $p-$type channel JFET. A clear transistor effect was measured under light exposure (\textit{active state}) while almost no current modulation was observed in the dark (\textit{freeze state}). The robustness of the non volatile photo-switch has been demonstrated during two days at room temperature and more than 140 minutes at $\SI{100}{\celsius}$ The high turn-on and turn-off times are the consequence of the thick, resistive, substrate used as the gate, the low light power density and the opaque metallic top contacts. They could be drastically reduced by simple changes to design of the device, opening the route to numerous applications in the field of power electronics, bio-electronics and quantum devices.

\section{Experimental Section}
\subsection{Diamond pn junction structure deposition}

The sample structure is depicted in Figure~\ref{Figure1}a. The $p-$type layer consists of a 400 nm boron-doped ([N$_{A}$-N$_{D}$]$=2\times 10^{17}$ cm$^{-3}$) diamond layer grown by microwave plasma assisted CVD, provided by DiamFab, on a $4\times 4\times0.5$  mm$^3$ Sumitomo High Pressure High Temperature (HPHT) Nitrogen-doped ($n-$type) diamond substrate, also referred as 1b diamond. Its nitrogen content is estimated to be in the order of $10^{19}$ cm$^{-3}$, mostly in single substitutional sites. The precise doping level was not measured but the sample exhibited the characteristic yellow tint of a high nitrogen content, as shown in the optical top view of the sample in Figure~\ref{Figure1}b. But, conversely, the $p-$type layer doping level was confirmed by capacitance-voltage measurements, not shown here, performed on metal oxide semiconductor capacitors test devices, fabricated on the same sample. 

\subsection{Fabrication of diamond transistor}

To lower the energy barrier between the top metal contacts and the $p-$type epilayer, a 200 nm highly boron-doped ([B]$>4\times 10^{20}$ cm$^{-3}$) diamond layer was selectively grown using an aluminium mask. A mesa etching was then performed down to the substrate by reactive ion etching with an oxygen plasma. This enabled the electric separation of each test devices on the sample surface from each other. The top metallic contacts were formed by a Titanium/Platinum/Gold (30 nm/20 nm/30 nm) stack deposited by electron beam evaporation and annealed for 30 min at $\SI{500}{\celsius}$ in high vacuum condition ($<10^{-7}$ mBar). An UV ozone treatment was performed on the surface before the deposition of 50 nm alumina layer by atomic layer deposition to passivate the whole surface, followed by post annealing in high vacuum condition ($<10^{-7}$ mBar) at $\SI{500}{\celsius}$ for 1 hour \cite{Pham2018}. The top drain and source contact consists of a Titanium/Platinum/Gold (30 nm/20 nm/30 nm) stack, deposited by electron beam evaporation on the 50 nm alumina layer. No lithography step has been performed after the passivation oxide deposition. Also the metal oxide metal stack thus formed is  electrically broken at the time of the first measurement by applying a few volts bias voltage. The detailed process is given in Ref. \cite{Masante2021}. The back contact between the substrate and the gold layer on the alumina plate was done thanks to silver paste.  

\subsection{Electrical characterization of the device}

The PN diode electrical characterization was performed using a Keithley 2612. The JFET electrical characterization was performed using a Modulab Solartron including the low current module. The electrical contacts were obtained using probes moved by a piezoelectric system with a LINKAM system to control temperature under vacuum. The microscope white LED light beam was used to optically excite the sample from the top. The use of a nitrogen-doped substrate as $n-$type layer is convenient and greatly simplifies the fabrication process. However it also causes a very large serial resistance $R_{Sub}$ due to its thickness (0.5 mm) and large resistivity. Also the silver paste back contact is probably of poor quality. Capacitance measurements are prohibited by the low cut-off frequency induced by the serial resistance so we only relied on static electrical measurements. This simple design was however found to be well suited to demonstrate this proof of concept.

\acknowledgements
The authors are thankful to David Eon (Univ. Grenoble Alpes, CNRS, Institut N\'eel, France), Etienne Gheeraert (Univ. Grenoble Alpes, CNRS, Institut N\'eel, France), Juliette Letellier (DiamFab, France), Khaled Driche  (DiamFab, France), Hitoshi Umezawa (National Institute of Advanced Industrial Science and Technology, Japan) and Robert Nemanich (Arizona State University, USA) for fruitful discussions.

\end{document}